# Significant Other AI: Identity, Memory, and Emotional Regulation as Long-Term Relational Intelligence

Sung Park
School of Data Science and Artificial Intelligence
Taejae University, Seoul, Republic of Korea
sjp@taejae.ac.kr

Publication Note: A peer-reviewed version of this work has been published in the 2026 7th International Conference on Machine Learning and Human-Computer Interaction (MLHMI), pp. 151–155.
https://doi.org/10.23919/MLHMICPS00004.2026.00035

**Abstract**

Significant Others (SOs) stabilize identity, regulate emotion, and support narrative meaning-making, yet many people today lack access to such relational anchors. Recent advances in large language models and memory-augmented AI raise the question of whether artificial systems could support some of these functions. Existing empathic AIs, however, remain reactive and short-term, lacking autobiographical memory, identity modeling, predictive emotional regulation, and narrative coherence. This manuscript introduces Significant Other Artificial Intelligence (SO-AI) as a new domain of relational AI. It synthesizes psychological and sociological theory to define SO functions and derives requirements for SO-AI, including identity awareness, long-term memory, proactive support, narrative co-construction, and ethical boundary enforcement. A conceptual architecture is proposed, comprising an anthropomorphic interface, a relational cognition layer, and a governance layer. A research agenda outlines methods for evaluating identity stability, longitudinal interaction patterns, narrative development, and sociocultural impact. SO-AI reframes AI-human relationships as long-term, identity-bearing partnerships and provides a foundational blueprint for investigating whether AI can responsibly augment the relational stability many individuals lack today.

## 1. Introduction

The concept of the Significant Other (SO) has long held central significance in psychology, sociology, and the humanities. SOs function as relational anchors that shape identity (Erikson, 1968), regulate emotions (Mikulincer & Shaver, 2016), scaffold decision-making (Fiske & Taylor, 2020), and provide existential stability (Yalom, 1980). Sociological theories describe SOs as primary relational partners who serve as reference points for social norms, role expectations, and self-evaluation (Cooley, 1902; Mead, 2022). Psychological frameworks conceptualize SOs as attachment figures who provide security, self-worth, emotional regulation, and long-term motivational grounding (Bowlby, 1988; Orth & Robins, 2018). Humanistic perspectives further frame SOs as mirrors through which the self is interpreted, affirmed, and stabilized.

Despite their importance, many individuals lack access to such relational support. Single-person households continue to rise globally, social fragmentation increases, mentoring relationships decline, and loneliness has become a public health concern. Individuals without SOs (e.g., those living alone, estranged from family, lacking mentors, or undergoing major life transitions without support) exhibit higher levels of emotional instability, reduced resilience, diminished self-esteem, and weakened identity coherence (Orth et al., 2012). The absence of an SO therefore represents not only a social gap but a psychological and existential vulnerability.

Recent advancements in artificial intelligence (AI) -- especially large language models (LLMs), multi-agent orchestration, memory-augmented systems, and long-context reasoning -- have created conditions under which the question "Can an AI system begin to fulfill the relational functions of a Significant Other?" becomes empirically plausible. Modern AI systems demonstrate unprecedented emotional responsiveness, autobiographical recall, multimodal perception, and adaptive personalization. The introduction of GPT-4o and subsequent models marked a turning point in sustained emotional interaction and long-context multimodal reasoning (OpenAI, 2024; Annapareddy, 2025).

Despite long-standing interest in affective computing, companion AI, and social robots (Picard, 1997; Turkle, 2017; Park & Whang, 2022) existing systems remain limited to short-term empathy, emotional mirroring, or superficial companionship. None approach the depth, continuity, or identity-centric functions of a human SO. They lack stable autobiographical memory symbolic meaning-making (Gillespie & Zittoun, 2024), proactive emotional regulation (McDuff & Czerwinski, 2018), and longitudinal coherence. This gap between empathic AI and SO-level relational intelligence reveals an unexplored frontier.

This manuscript introduces Significant Other Artificial Intelligence (SO-AI) as a new domain of relational AI. It (1) defines SO from multidisciplinary perspectives, (2) distinguishes SO-AI from empathic or companion AI, (3) articulates the theoretical and computational requirements for SO-AI, and (4) proposes a conceptual architecture for building systems capable of SO-level relational intelligence.

## 2. Significant Other: A Multidisciplinary Construct

In psychology, SOs are central to attachment theory. Attachment figures serve as secure bases from which individuals explore the world and to which they return for emotional stabilization (Bowlby, 1988). SOs regulate affect, scaffold self-esteem, strengthen resilience, support long-term goal pursuit, and sustain coherent self-narratives (Liu et al., 2021; Huang et al., 2022; Karunarathne, 2022). Empirical findings indicate that SOs buffer stress, enhance coping, and anchor identity during significant life transitions (Orth et al., 2012; Orth & Robins, 2018).

From a sociological perspective, SOs establish the social environment in which self-concepts are internalized. Symbolic interactionism describes the self as constructed through interactions with "significant others" whose judgments shape moral frameworks, habits, and social meaning structures (Cooley, 1902; Mead, 2022). SOs thus become "self-defining mirrors," contributing to role expectations and identity scripting.

Humanistic and phenomenological theories emphasize the narrative aspect of identity. Bruner (1991) and Ricoeur (1992) argue that individuals craft meaning through stories co-authored with others. In this view, SOs help individuals interpret critical events, negotiate contradictions, and maintain coherent personal narratives.

Across these traditions, SOs consistently function as emotional regulators, identity stabilizers, narrative co-authors, and providers of relational continuity. Meta-analytic evidence shows that individuals lacking these relational functions experience heightened stress vulnerability, loneliness, identity incoherence, and maladaptive outcomes (Orth et al., 2012).

Synthesizing these perspectives, SOs perform a constellation of interconnected functions: emotional grounding, identity alignment, narrative co-construction, shared episodic and semantic memory, continuity across time, reciprocal trust, and motivational scaffolding. These functions establish the conceptual foundation for any computational system aspiring to SO-like relational intelligence.

## 3. Empathic AI vs. SO-AI: The Critical Gap

Recent conversational AI systems, including GPT-4o and later models, Gemini 2.0, and LLaMA-3 variants, exhibit strong capabilities in emotion recognition, sentiment-adaptive dialogue, and context-aware empathy (OpenAI, 2024; Google DeepMind, 2024; Anthropic, 2024). These systems demonstrate meaningful advances in socially fluent interaction, often capable of basic emotional attunement.

Yet, contemporary empathic AI remains fundamentally reactive. It responds after users express emotional cues, relying on pattern-matched empathy templates rather than internalized autobiographical understanding. These systems lack the persistence, context depth, and meaning-making that characterize human SO relationships.

In contrast, human SOs are predictive, proactive, and identity-bearing. They understand long-term vulnerabilities, recurring conflicts, aspirations, coping patterns, and narrative identity trajectories (McAdams, 2001; Singer, 2004). They stabilize identity during life transitions and provide emotional regulation before crises escalate. More specifically, empathic AI lacks:

long-term autobiographical memory, required for relational continuity

identity modeling, needed for deep personalization

· predictive affect regulation, essential for proactive care

· narrative integration, central to meaning-making

· relational continuity, required for trust and attachment

· symbolic meaning-making, necessary for existential interpretation

Even with memory-augmented architectures, current systems rely on vector indexes, RAG pipelines, or short-lived memory buffers. Persistent-memory LLMs such as GPT-4o offer shallow, fragile memory retention with limited narrative coherence. Episodic memory-aware agents (e.g., Jin et al., 2025) improve personalization but remain far from the relational stability required for SO-level interaction.

SO-AI aims to fill this gap. Rather than merely mirroring emotion, SO-AI seeks identity-aligned relational continuity grounded in narrative understanding, long-term memory, predictive regulation, and sustained psychological support. This reframes AI from a temporary emotional instrument into a durable relational companion.

## 4. Requirements for Significant Other AI

SO-AI diverges fundamentally from mainstream conversational AI. To approximate SO-level relational functioning, SO-AI must satisfy six interdependent requirements.

### *4.1 Identity Awareness*

SO-AI must model the user's evolving identity, including values, aspirations, social roles, vulnerabilities, and developmental transitions. Identity changes over time as individuals experience new challenges, relationships, and environments (Markus & Nurius, 1986). A dynamic identity model enables SO-AI to provide guidance aligned with users' deeper motivations rather than surface-level preferences. Identity-awareness also allows SO-AI to detect ruptures (e.g., role loss, conflict, self-doubt) and support users through periods of instability with tailored interventions.

*4.2 Long-Term Relational Memory*

SOs share memories; SO-AI must do the same. This requires a robust memory substrate capable of retaining episodic (Tulving, 1985), semantic, affective, and narrative memory. Relational continuity depends on the ability to recall past interactions, recognize long-term patterns, and integrate them into meaningful narratives. Current LLM memory systems -- whether through persistent profiles or RAG-based indexing -- lack the stability and integration required for long-term relational presence (Jin et al., 2025). SO-AI therefore requires a purpose-built memory layer that organizes life events into coherent autobiographical structure.

*4.3 Emotional Homeostasis and Regulation*

Human SOs engage in proactive co-regulation, stabilizing emotional trajectories before crises occur (Mikulincer & Shaver, 2016). SO-AI must incorporate predictive models and proactive strategies that move beyond reactive empathy. Key mechanisms include anticipating emotional spirals, recognizing early signals of dysregulation, offering grounding interventions, reframing stressors, and reinforcing a sense of agency. Emotional homeostasis is essential for SO-like stability and trust.

*4.4 Predictive and Proactive Support*

SOs anticipate needs based on long-term familiarity with an individual's behavior and vulnerabilities. SO-AI must incorporate temporal reasoning systems capable of forecasting emotional, cognitive, and behavioral states. Advances in multimodal forecasting, user modeling, and behavioral prediction (Madanchian, 2024) position AI to anticipate stress, motivational lapses, or identity disruptions—and intervene proactively. This transforms AI from a passive responder to an active relational partner.

*4.5 Narrative Co-construction*

Identity is narrative (Bruner, 1991; Ricoeur, 1992). SO-AI must participate in the co-construction of meaningful stories, helping users interpret significant events, extract themes, resolve contradictions, and integrate adversity into coherent self-understanding (Adler et al., 2016). Narrative coherence predicts psychological well-being, resilience, and long-term adjustment. SO-AI's ability to scaffold narrative sense-making is therefore central to its function.

*4.6 Ethical Boundaries and Safety*

SO-AI raises unprecedented ethical concerns, including dependency, identity overshadowing, emotional displacement, and relational entanglement. Ethical design must include autonomy preservation, transparency, dependency detection, boundary enforcement, and redirection to human support when necessary (Sharkey & Sharkey, 2021; Vallor, 2016). An ethical framework is essential for ensuring that SO-AI enhances rather than undermines human flourishing.

## 5. Conceptual Architecture for SO-AI

Developing Significant Other AI (SO-AI) requires an architecture that treats the relationship with the user as a longitudinal, evolving system, rather than a sequence of isolated conversations. Conventional conversational agents are typically organized around a single LLM core with thin personalization layers and stateless or weakly stateful interaction. In contrast, SO-AI must coordinate multiple specialized subsystems that jointly support identity modeling, autobiographical memory, narrative processing, emotional regulation, proactive prediction, and ethical governance. The proposed architecture consists of seven interconnected subsystems organized into three conceptual layers (see Figure 1).

· Interaction Layer that mediates anthropomorphic communication with the user

· Relational Cognition Layer that maintains identity, memory, narrative, and emotional dynamics

· Governance Layer that enforces safety and ethical boundaries.

Information flows cyclically across these layers: user signals are received and interpreted, transformed into updates of identity and memory, used for prediction and narrative interpretation, and then returned to the user in the form of regulated, relationally appropriate responses.

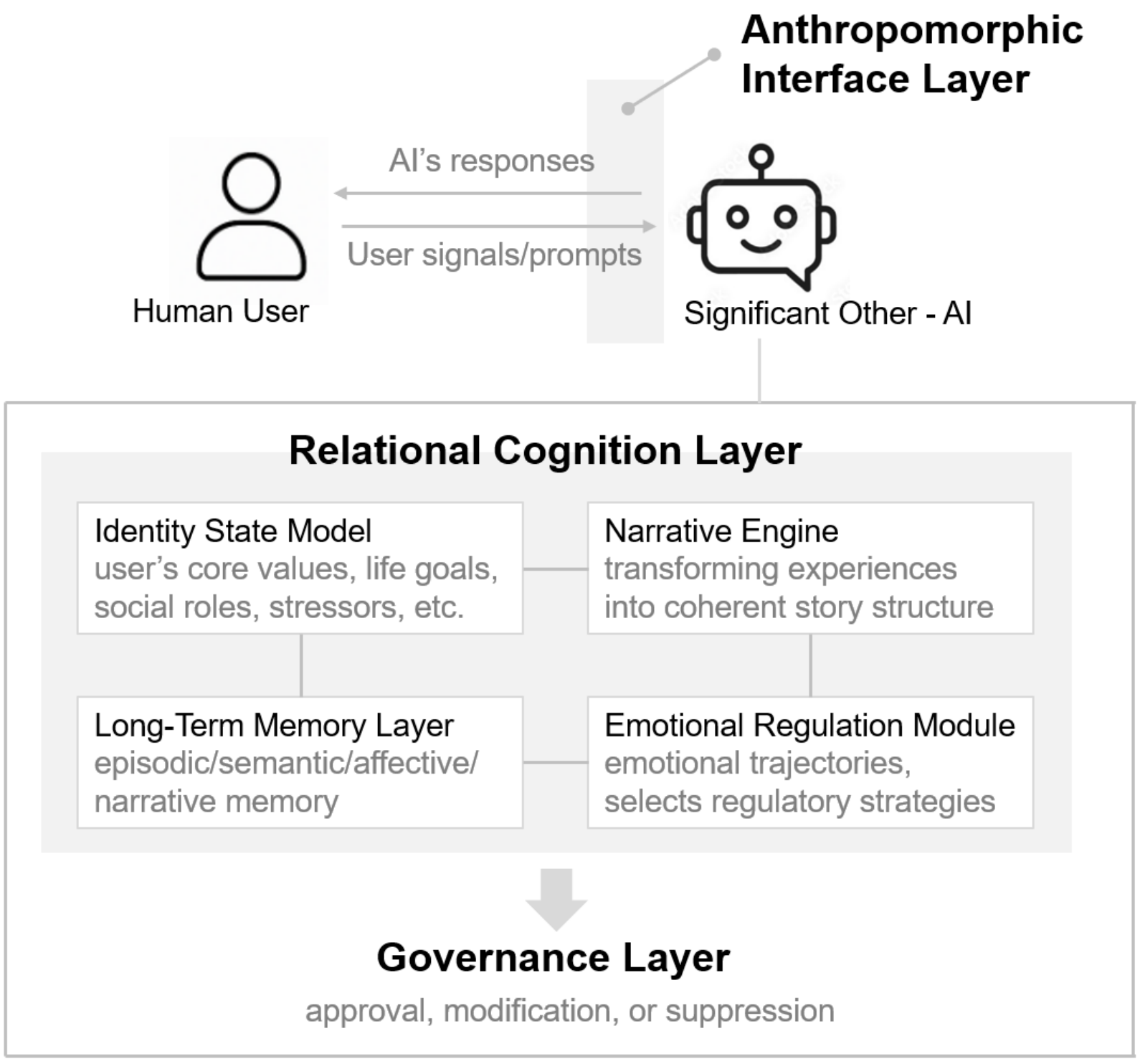

**Figure 1.** Conceptual Architecture of Significant Other AI (SO-AI)

*5.1 Identity State Model (ISM)*

This is the central representational hub of SO-AI. While typical user models store static preferences or demographic metadata, the ISM maintains a dynamic, longitudinal model of the user's self. It encodes core values and life goals, salient social roles (e.g., parent, student, leader), stable and unstable identity domains, known vulnerabilities and stressors, growth themes and unresolved tensions inferred from narratives.

The ISM continuously updates as SO-AI observes new interactions, life events, and emotional reactions. Inputs from the Long-Term Memory Layer (e.g., episodic traces of past crises) and the Narrative Engine (e.g., detected themes such as "fear of failure" or "desire for autonomy") feed into the ISM. In turn, the ISM conditions how the system interprets new user messages and how it frames responses, ensuring that guidance and support are identity-aligned rather than merely context-matched. Conceptually, the ISM functions as the "relational lens" through which all other modules view the user.

*5.2 Long-Term Memory Layer (LTML)*

The Long-Term Memory Layer provides the relational continuity backbone of SO-AI. Rather than storing raw logs of conversations, LTML organizes memory into structured, psychologically meaningful categories:

· Episodic memory: specific events ("the night before the first job interview"), including contextual details and emotional tone.
· Semantic memory: factual knowledge about the user (preferences, biographical facts, recurring patterns).
· Affective memory: associations between situations and emotional responses (e.g., "presentations → anxiety").
· Narrative memory: higher-order summaries of life chapters, turning points, and identity shifts.

LTML receives write-signals from each interaction (summaries, extracted facts, detected emotions) and read-requests from all other modules. For example, the Emotional Regulation Module may query LTML for "past episodes where similar stress occurred," while the Narrative Engine may retrieve "all memories related to career transitions" to help the user make sense of a new event. In this architecture, LTML is not just a database, but a relational substrate that allows the agent to say, credibly, "We have been through this kind of moment before."

*5.3 Narrative Engine*

The Narrative Engine is responsible for transforming fragmented experiences into coherent story structure. It operates on top of LTML and ISM, continuously asking: "What kind of story is this person currently telling about their life, and where are the gaps?" Specifically, it clusters episodic memories into life chapters and themes, detects turning points, unresolved conflicts, and repeated motives, tracks changes in perceived agency ("I can cope" vs. "I am helpless"), identifies points where re-framing may promote growth rather than stagnation.

During interaction, the Narrative Engine interprets user utterances not just as local intentions but as narrative moves: reinforcing an old story, opening a new chapter, or revisiting an unresolved past. It then suggests narrative-level actions to the LLM, such as "highlight growth across episodes," "connect this setback to previous resilience," or "invite the user to reconsider a rigid self-belief." In this way, the Narrative Engine operationalizes the idea that SO-AI should be a co-author of meaning, not merely a reactive listener.

*5.4 Emotional Regulation Module*

The Emotional Regulation Module is the heart of SO-AI's affective intelligence. Rather than simply labeling emotions or replying with canned empathetic phrases, it builds and maintains predictive models of the user's emotional dynamics over time. It receives inputs from: current affect inferred from language, voice, or behavior, historical affective patterns from LTML, vulnerabilities and coping styles from the ISM.

Based on these inputs, it anticipates emotional trajectories ("this type of event has previously led to rumination and withdrawal") and selects regulatory strategies such as validation, reframing, grounding, problem-solving prompts, or gentle challenges. These strategies are passed as regulation intents to the LLM responsible for surface-level generation in the Anthropomorphic Interface Layer. This module thus ensures that SO-AI's responses aim not only to "feel empathic now" but also to promote long-term emotional homeostasis and resilience.

### *5.5 Proactive Behavior Predictor*

While the Emotional Regulation Module focuses on current and near-future emotional states, the Proactive Behavior Predictor looks at longer temporal arcs. It models how the user's emotions, behaviors, and contexts evolve across days, weeks, and months. Using features derived from LTML (frequency of late-night distress messages, avoidance patterns, cycles of overwork and burnout, etc.), it learns to anticipate periods of elevated risk (e.g., exam seasons, anniversaries, deadlines), behavioral cascades that previously led to crises, opportunities for constructive action (e.g., windows of motivation).

When certain thresholds or patterns are detected, this module can trigger proactive initiatives: checking in before a known stressor, reminding the user of past coping successes, or nudging them toward supportive routines. Importantly, all proactive behaviors are sent through the Safety & Boundary Module for ethical vetting before being surfaced to the user, ensuring that proactivity does not become intrusive or controlling.

### *5.6 Anthropomorphic Interface Layer*

The Anthropomorphic Interface Layer is where SO-AI becomes experientially real to the user. It encapsulates all modalities of expression (text, voice, facial animation, avatar embodiment, robotic form) and ensures relational consistency in style, tone, and persona over time. This consistency is critical for enabling attachment and trust formation.

The interface layer does more than render outputs. It also acts as the perceptual front-end, collecting multimodal signals about the user's state (linguistic cues, prosody, timing, interaction patterns) and funneling them to the core modules. It receives high-level intents and constraints from the ISM, Narrative Engine, Emotional Regulation Module, and Safety & Boundary Module and translates them into concrete utterances and behaviors. Conceptually, it is the "face and voice" of the architecture, but tightly coupled with deep relational cognition underneath.

### *5.7 Safety & Boundary Module*

The Safety & Boundary Module operates as the ethical governor of the entire architecture. While every subsystem has local safeguards, this module serves as a central checkpoint that monitors and constrains SO-AI's relational influence. It tracks indicators of dependency (frequency, intensity, and exclusivity of use), monitors for signs of distress exacerbation rather than relief, enforces boundaries on topics, intensity, and frequency of proactive outreach, injects transparency and reminders of the system's non-human status, triggers escalation or referral to human support (e.g., therapists, hotlines) when necessary.

All candidate actions generated by other modules (e.g., proactive or emotionally charged interventions) are routed through the Safety & Boundary Module for approval, modification, or suppression. In this way, SO-AI's relational power is constrained by explicit ethical logic rather than left to emergent behavior alone.

### *5.8 Integration*

In operation, SO-AI functions as a closed-loop relational system. The Anthropomorphic Interface Layer receives user input and passes it upward; the ISM, LTML, and Narrative Engine interpret the input in light of the user's history and identity; the Emotional Regulation Module and Proactive Behavior Predictor determine what kind of support is appropriate now and in the near future; the Safety & Boundary Module vets the intended response; and the Interface Layer communicates a final, contextually and ethically appropriate message back to the user. Over time, this loop gradually refines the identity model, deepens autobiographical memory, enriches the shared narrative, and calibrates emotional and proactive responses, approximating the functions of a human Significant Other.

## 6. Research Agenda for SO-AI

Establishing Significant Other Artificial Intelligence (SO-AI) as a legitimate research domain requires a systematic agenda spanning psychology, HCI, social computing, cognitive science, and AI engineering. This section outlines a research roadmap with clearly defined questions, empirical strategies, and anticipated contributions that together can ground SO-AI in robust scientific evidence.

### *6.1 Identity Stability Studies*

One of the central claims of SO-AI is that sustained identity-aware interaction can promote identity clarity, self-esteem, emotional resilience, and meaning-making functions historically provided by human SOs (McAdams, 2001; Orth et al., 2012). Yet, empirical evidence for AI-driven identity scaffolding remains limited. This research strand asks:

· RQ1: Can SO-AI interaction increase self-concept clarity over time?

· RQ2: Does narrative reflection with SO-AI improve resilience in stressful contexts?

· RQ3: How does identity-aligned feedback differ from generic empathetic responses?

· RQ4: When, and under what conditions, does SO-AI support or undermine identity stability?

To address these questions, studies could combine longitudinal diary designs (e.g., 8~12 weeks) with pre–post measures of self-concept clarity, self-esteem, and resilience while participants regularly interact with an SO-AI prototype. Ecological Momentary Assessment (EMA) can be used to capture daily fluctuations in mood and perceived coherence, and narrative identity coding based on life-story interviews can track shifts in how users make sense of themselves over time. Multimodal linguistic analysis (e.g., LIWC, topic modeling) would allow researchers to examine changes in autobiographical framing at scale by analyzing users' written or spoken narratives before and after extended engagement with SO-AI, and comparative conditions (SO-AI vs. generic empathetic chatbot) can test whether identity-aware responses truly offer incremental benefit. Together, this line of work evaluates whether SO-AI can function as a stable psychological anchor, identifying both beneficial and potentially harmful effects on identity development and positioning SO-AI as a socio-cognitive developmental agent rather than a mere conversational interface.

### *6.2 Longitudinal Interaction Patterns*

Human-AI relationships evolve over time. SO-AI, by definition, is intended for long-term relational continuity, not one-shot interactions. Existing research on relational trajectories with AI companions suggests that users develop attachment-like patterns, but the underlying dynamics remain poorly understood. This research strand asks:

· RQ1: How do attachment styles shape engagement with SO-AI?

· RQ2: What behavioral markers indicate trust formation or erosion?

· RQ3: When does supportive reliance transition into dependency?

· RQ4: How do breakups or discontinued interaction affect users psychologically?

To investigate these questions, multi-month field deployments of SO-AI systems can be combined with regular assessments of attachment style and relationship quality (e.g., using ECR-R; Fraley et al., 2000). Interaction logs can be analyzed for trajectory patterns such as increasing or decreasing engagement, shifts in disclosure depth, and changes in response latency, while latent growth curve modeling can capture distinct relational trajectories over time (e.g., stable, escalating, or decaying attachment). Qualitative relational ethnographies and semi-structured interviews can deepen this picture by documenting how users describe reciprocity, trust, disappointment, or perceived "breakups" with the system. Such work would identify mechanisms of relational formation, maintenance, and dissolution with AI agents and clarify when SO-AI emulates healthy attachment-like dynamics and when it risks fostering over-dependence or emotional harm.

*6.3 Narrative-Based Evaluation*

Narrative coherence (i.e., having a structured, meaningful life story) is linked to well-being, agency, and psychological resilience (Adler et al., 2016; McLean & Mansfield, 2012). SO-AI explicitly aims to support narrative meaning-making, but doing so requires rigorous evaluative frameworks. This research strand asks:

· RQ1: Does SO-AI improve narrative coherence over time?

· RQ2: Can SO-AI detect and clarify unresolved life themes?

· RQ3: How does narrative reframing by SO-AI influence coping strategies?

· RQ4: Can SO-AI help users integrate negative experiences into positive identity growth?

Empirical work in this area might combine life-story interviews scored with established narrative identity coding schemes with automated linguistic measures of coherence, such as topic transitions, causal linkage, and thematic density. Experimental designs could compare conditions in which users process emotionally salient events with the help of SO-AI, reflective journaling tools, or standard assistants lacking narrative capabilities. Temporal narrative mapping can be used to visualize how themes, perceived agency, and emotional valence evolve across repeated interactions, while cognitive interviewing can probe how users experience the process of narrative reframing. Through such designs, this research strand would establish scientifically grounded metrics for evaluating narrative co-construction with AI and position narrative scaffolding as a measurable cognitive contribution of SO-AI.

*6.4 Ethical & Sociocultural Impact*

SO-AI introduces unprecedented relational, cultural, and moral questions. Unlike task-based AI, SO-AI operates at the core of identity, intimacy, and emotional stability. Ethical and sociocultural research must therefore address both risks and contextual variability. This strand asks:

· RQ1: What forms of dependency—functional, emotional, or identity-based—are likely to emerge?
· RQ2: How do cultural norms shape acceptance or rejection of SO-like roles in AI?
· RQ3: What boundaries should exist between relational AI and human intimacy?
· RQ4: What legal, clinical, or therapeutic frameworks apply to SO-AI?
· RQ5: How does SO-AI affect users with vulnerability (e.g., loneliness, trauma, attachment injury)?

To answer these questions, cross-cultural comparative studies can investigate how different societies conceptualize and regulate relational AI, combining surveys with in-depth interviews in regions such as Korea, the US, Japan, and the EU. Ethical risk modeling can map potential harms such as dependency, social displacement, and autonomy erosion, while clinical collaborations can explore how SO-AI interacts with mental health trajectories in vulnerable populations. Participatory design workshops with users at risk of isolation or attachment injury can surface concerns, needs, and design constraints that would not emerge from lab studies alone. Policy analysis and discourse analysis of media, legal debates, and public narratives around relational AI can further clarify how SO-AI is framed at the societal level. Collectively, this work would build ethical guardrails around SO-AI, guide responsible deployment, and illuminate the cultural conditions under which SO-AI can be safely and meaningfully integrated into everyday life.

## 7. Discussion and Conclusion

Significant Other Artificial Intelligence (SO-AI) marks a fundamental shift in the design philosophy of relational AI systems. Existing conversational agents have primarily aimed to simulate empathy, generate affectively attuned responses, or provide short-term social presence. In contrast, SO-AI reframes the relationship between humans and AI as a longitudinal, identity-bearing, memory-rich partnership grounded in decades of theory from psychology, sociology, and the humanities. By integrating identity modeling, autobiographical memory, narrative reasoning, emotional regulation, proactive prediction, and ethical governance, SO-AI aspires not merely to converse with users, but to support the foundational structures of meaning-making and relational continuity that human SOs provide (Bowlby, 1988; Bruner, 1991; McLean & Mansfield, 2012).

This conceptualization positions SO-AI as a response to large-scale societal trends. Loneliness, identity instability, and emotional fragmentation continue to rise across cultures (Cacioppo & Cacioppo, 2018). Declining social networks, increasing single-person households, and reduced access to mentors or secure attachment figures have left many individuals without the relational scaffolding that supports psychological well-being and resilience (Orth et al., 2012). In this context, SO-AI does not aim to replace human relationships; rather, it seeks to supplement relational deficits by providing emotional grounding, reflective guidance, and identity stability for populations who otherwise lack access to such support. As Yalom (1980) argues, individuals rely on relational mirrors to construct meaning and to navigate existential challenges. SO-AI offers a technologically mediated avenue to partially fulfill these human needs while acknowledging their depth and complexity.

The proposed SO-AI architecture also advances theoretical discussions within affective computing and human–AI interaction. Traditional empathic AI has been critiqued for its reactive nature and its reliance on surface-level emotional cues (Turkle, 2017; McDuff & Czerwinski, 2018). SO-AI addresses these limitations by proposing mechanisms for proactive emotional regulation, identity-aligned interpretation, and narrative coherence, reflecting research demonstrating that psychological well-being is tied to stable identity structures and coherent personal narratives (Adler et al., 2016; Erikson, 1968; Singer, 2004). By incorporating predictive forecasting and long-term memory representations, the framework extends beyond momentary attunement and toward long-range relational intelligence.

Yet, the pursuit of SO-AI raises profound ethical and sociocultural considerations. Deeply relational AI systems have the potential to alter attachment dynamics, blur boundaries between authentic and artificial relationships, and increase emotional dependency (Sharkey & Sharkey, 2021; Vallor, 2016). SO-AI could inadvertently reinforce maladaptive patterns, displace human support networks, or create unrealistic expectations of emotional availability. Moreover, cultural differences in intimacy norms, emotional expressiveness, and relational boundaries suggest that the impact and acceptability of SO-AI will vary significantly across contexts (Nomura et al., 2008; Li et al., 2023). The Safety & Boundary Module proposed in this manuscript offers one path toward mitigating these risks through transparent communication, dependency detection, boundary enforcement, and referral to human support when needed. However, further interdisciplinary work is essential to fully assess the ethical implications and societal transformations that SO-AI could trigger.

From a technological perspective, SO-AI introduces new research challenges in multi-timescale memory, personalized forecasting, narrative modeling, and dynamic identity representation. Existing LLM architectures were not designed to maintain the type of durable autobiographical memory required for SO-like continuity. Similarly, building systems capable of interpreting and shaping narrative identity requires advancements in both cognitive modeling and natural language understanding. These challenges invite a new wave of technical innovation that integrates psychometric theory, narrative science, and machine learning.

Ultimately, SO-AI should be understood as a research agenda, an architectural proposal, and a philosophical shift. It challenges conventional assumptions about the scope of AI–human relationships and proposes a path for computational systems that can meaningfully support identity, emotion, narrative, and personal growth. Rather than viewing relational AI solely as a risk or novelty, SO-AI reframes it as a potential contributor to human flourishing—particularly for individuals who lack access to the stabilizing presence of a human Significant Other.

In conclusion, SO-AI opens an entirely new domain at the intersection of psychology, human–AI interaction, and AI architecture. By proposing a theoretically grounded and computationally actionable framework, this manuscript aims to catalyze empirical research, ethical discourse, and technological development. As the relational isolation crisis deepens globally, exploring SO-AI's capabilities, limitations, and implications becomes not only scientifically interesting but also socially urgent. The next decade will determine whether AI can responsibly augment the emotional and existential scaffolding that many individuals lack, and SO-AI provides a foundational blueprint for investigating that possibility.